\documentclass[aip, jmp, reprint]{revtex4-2}
\usepackage[utf8]{inputenc}
\usepackage{hyperref}

\usepackage{xcolor}
\usepackage{graphicx}
\usepackage{dcolumn}
\usepackage{bm}

\usepackage[utf8]{inputenc}
\usepackage[T1]{fontenc}
\usepackage{mathptmx}
\usepackage{etoolbox}

\makeatletter
\def\@email#1#2{%
 \endgroup
 \patchcmd{\titleblock@produce}
  {\frontmatter@RRAPformat}
  {\frontmatter@RRAPformat{\produce@RRAP{*#1\href{mailto:#2}{#2}}}\frontmatter@RRAPformat}
  {}{}
}%
\makeatother
\begin{document}

\preprint{AIP/123-QED}

\title{Introduction to Focus Issue: Topics in Nonlinear Science
}
\author{Elizabeth Bradley}
 \email{lizb@colorado.edu}
 \affiliation{Department of Computer Science, University of Colorado, Boulder, Colorado 80309-0430, USA; Santa Fe Institute, Santa Fe, New Mexico 87501, USA.}
\author{Adilson E. Motter}
\affiliation{Department of Physics and Astronomy,  Center for Network Dynamics, Department of Engineering Sciences and Applied Mathematics, and Northwestern Institute on Complex Systems, Northwestern University, Evanston, Illinois 60208, USA.}
\author{Louis M. Pecora}
\affiliation{Institute for Research in Electronics and Applied Physics (IREAP), University of Maryland, College Park, Maryland 20742, USA.}


\begin{abstract}
Nonlinear science has evolved significantly over the 35 years since the launch of the journal {\it Chaos}. This Focus Issue, dedicated to the 80th Birthday of its founding editor-in-chief,
David K. Campbell, brings together a selection of contributions on influential topics, many of which were advanced by Campbell's own research program and leadership role. The topics include new phenomena and method development in the realms of network dynamics, machine learning, quantum and material systems, chaos and fractals, localized states, and living systems, with a good balance of literature review, original contributions, and perspectives for future research.
\end{abstract}

\maketitle

\begin{quotation}

Writing about the future of nonlinear science nearly 40 years ago, Campbell asserted\cite{campbell1987nonlinear},
``To ensure the long-term success of nonlinear science, it is crucial to train young researchers in the paradigms of nonlinearity. Also, interdisciplinary networks must be fostered that consist of scholars who are firmly based in individual disciplines but are aware of, and eager to understand, developments in other fields.'' Through his leadership and deep commitment to these principles, he helped build not only a thriving field but also a vibrant global community.
This Focus Issue looks both back and forward at the state of the field, reaffirming that nonlinear science is, in Campbell's words,  ``broadly interdisciplinary, intellectually unfettered and demanding, and---very importantly---fun.''
\end{quotation}

\section{\label{sec:level1}Introduction}
This Focus Issue, which was organized in honor of a milestone birthday of David K. Campbell, the founding editor-in-chief of this journal, contains 37 invited papers involving a total of 116 authors, at least 23 of whom were official or unofficial mentees of Campbell. Notable in this selection of papers is not only the broad range of topics that are covered, but also the number and tone of the dedications and expressions of thanks and respect that they contain---as well as the specific instances that they offer of Campbell's central and lasting influence upon the field of nonlinear science.

Campbell’s own research contributions have ranged from fundamental results on the transverse field Ising model, the basis for quantum annealing architectures, and the double-well equation \cite{bertalan2024transformations, bouche2024zeptonewton} to the identification of the main paradigms of the field: solitons and coherent structures, deterministic chaos and fractals, complex configurations and pattern formation, and adaptive nonlinear systems \cite{ecke2024center}. This now-familiar taxonomy served as an effective scaffolding for the field: both its overarching principles and its applications. 
Campbell made major scientific contributions in many of those four areas, notably in elucidating the conditions and mechanisms underlying solitary waves and discrete breathers \cite{lee2024complexified}, as well as in pioneering the nonlinear dynamics of polyacetylene \cite{savin2024stabilization}.
In a long series of papers with many different co-authors---including one in this Focus Issue written with two of his current Ph.D. students \cite{karve2024periodic}---he explored various aspects of the Fermi-Pasta-Ulam problem, which broke open the idea that one could do mathematics with computers (“experimental mathematics”). In the past decade, Campbell played a key role in unearthing the “hidden figure” in the FPU story: Mary Tsingou Menzel, who was a partner in this work and whose name is now part of the identifier of that problem. 

These are only a few examples from a long and rich career. As Marc Timme writes, ``By often leaving well-trodden paths and pointing to novel perspectives, [Campbell] has left his personal mark in the field'' \cite{lee2024complexified}.  
Many of the papers in this Focus Issue were catalyzed by Campbell’s work, as exemplified by the acknowledgment: 
``motivated by the pioneering papers of David Campbell on nonlinear dynamics of molecular structures, here we have studied numerically the structure and dynamics of several types of hydrogen-bonded molecular chains placed inside capped or open-edge carbon nanotubes'' \cite{savin2024stabilization}. And he remains an active contributor to the field, as evidenced by the two papers in this Focus Issue on which he is a co-author \cite{karve2024periodic, bouche2024zeptonewton}, as well as the continued vitality of his research group at Boston University.

Campbell's formative role at the Los Alamos National Lab (LANL), where much of nonlinear science was born, began in 1974, when he joined the Lab as the first awardee of the J. Robert Oppenheimer Postdoctoral Fellowship. Over the next three decades, he rose through the ranks at LANL, playing roles in both new science and rich collaborations. These included the “Unstable Working Group,” which was formed to ponder how the fields of nonlinear dynamics and chaos were evolving. From these roots, the Center for Nonlinear Studies (CNLS) was formally created in 1980. Campbell directed this center on and off until 1993, building connections between people and providing creative guidance regarding both science and the navigation of administrative hurdles, such as the lack of office space (which was solved by renting a small adobe house in El Rancho, without prior permission from the accountants).

Robert Ecke's paper in this Focus Issue---from which much of the material in the previous paragraph is drawn---offers a detailed blow-by-blow chronicle of CNLS, complete with the names of everyone who was involved, either directly or indirectly, copies of posters from conferences and meetings, handwritten agendas for important meetings, early images of canonical results, and much more. Here, too, Campbell's profound influence shines through: ``Throughout [my time at the CNLS, from postdoc to Director], the inspiration and influence of David Campbell guided my way. Throughout its existence, CNLS owes much to the enduring legacy of David Campbell, who laid down the foundations and operating principles that have made it so successful... [he] was an inspiring leader, and I learned much under his tutelage. He, more than anyone, created the aura, environment, and spirit of CNLS.''\cite{ecke2024center}.

Campbell's mentoring has been particularly impactful for the dozens of junior scientists to whom it has been offered over the years, including the editors of this Focus Issue. These sentiments are expressed eloquently in the appendix of Ref.\ [\citenum{brooks2025opinion}], in which Mason Porter lauds Campbell as a ``great human being'' who has ``... continued looking out for me, as he has looked out for many others.'' Porter continues: ``... He has always been fair to me and he has been very good to me, as he has been to so many others... treating people fairly while simultaneously ensuring rigorous standards, and doing things the right way in scientific editing and reviewing. Some scientists consistently do things the right way, and David is one of them.''

\begin{figure}
\includegraphics[width=4in]{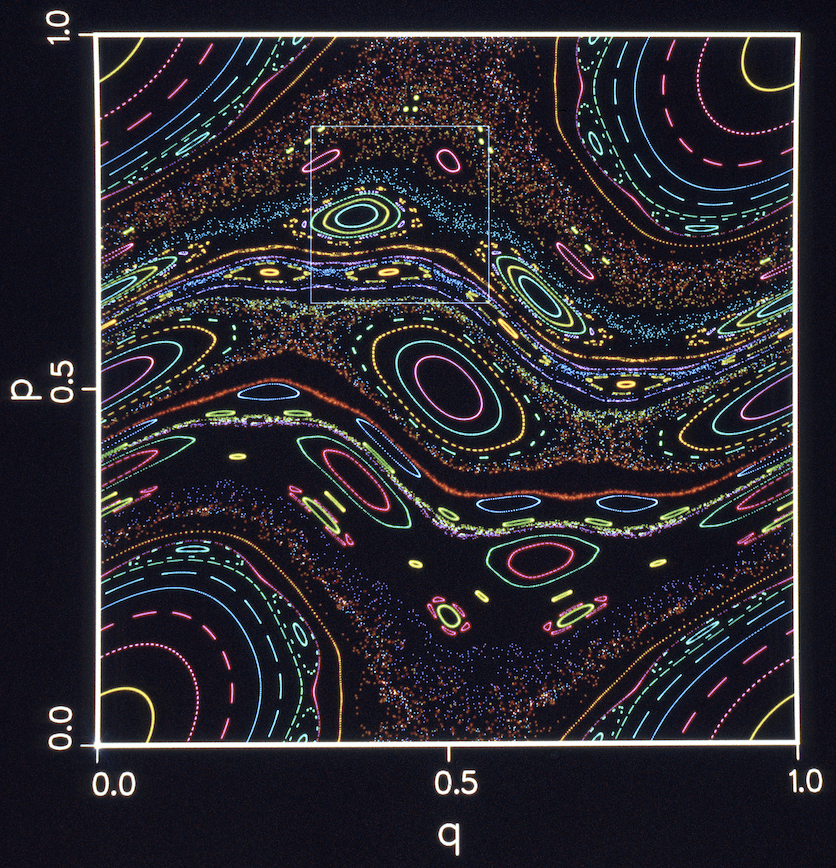}
\caption{\label{fig:fig1} 
Phase-space trajectories of a periodically kicked rotor, illustrating the coexistence of regular and chaotic regions in conservative systems. The coordinates represent the 
position and angular momentum in a stroboscopic map of the system. 
(Adapted from D. K. Campbell, Los Alamos Sci. Special Issue, 218–262 (1987)\cite{campbell1987nonlinear}).}
\end{figure}

One of us (Adilson Motter) had the distinct privilege of collaborating with Campbell on the Physics Today article\cite{motter2013chaos} commemorating the 50th anniversary of Lorenz’s landmark publication, “Deterministic Nonperiodic Flow”\cite{lorenz1963deterministic}. The experience was particularly rewarding, as Campbell brought not only a remarkable clarity of understanding of the field, but also deep insight into the individuals whose work shaped its development. It is largely thanks to his inspiring perspective that the article was later selected for inclusion in The Best Writing on Mathematics 2014 (Princeton University Press). This speaks not only to the depth of his scientific insight, but also to his singular ability to communicate the excitement and significance of nonlinear science to diverse audiences---an ability that was fittingly recognized with the 2010 APS Julius Edgar Lilienfeld Prize. A vivid example of this communicative gift, combined with his aesthetic sensibility, appears in his use of images to convey the intrinsic visual beauty of nonlinear science---as illustrated in Fig.\ 
\ref{fig:fig1}, which captures both the visual and conceptual elegance of chaotic systems with mixed phase spaces.

The origin story of the journal {\it Chaos} is intertwined with the history of the Soviet-American Chaos/XAOC conferences that were held on alternating sides of the Iron Curtain from 1989 to 1993. This workshop series led directly to the formation of the new AIP journal on nonlinear science for which this Focus Issue has been organized, 35 years later. Campbell’s article\cite{campbell2015pre} in the journal's 25th anniversary Focus Issue\cite{bradley2015introduction} recounts these matters in some detail (though not without downplaying his own role). In the Appendix, Kenneth Ford, the director of the American Institute of Physics from 1987 to 1993, covers some of this history as well, tracing the international impact and respect that the journal has attained, as well as the “smashing success” of the Chaos/XAOC conferences, directly back to Campbell's leadership.

\section{Papers in this issue}

The papers in this issue can be loosely divided into seven topical areas: networks, machine learning, quantum systems, materials, chaos and fractals, localized states, and living systems. These are in addition to Robert Ecke’s paper \cite{ecke2024center}, which is discussed above, as well as the two from Campbell and his students \cite{karve2024periodic, bouche2024zeptonewton} and a thought piece about data analysis from Bradley {\it et al.}\cite{bradley2024thoughtful}.

Various papers address problems related to complex networks and their dynamics.
The ability to control entire networks is highly desirable, but achieving this using only local information from a few nodes would be far more efficient. The paper by Haber {\it et al.}\cite{haber2024global} demonstrates that, for numerous networks, this is possible through a closed-form, non-iterative solution that relies on small state information neighborhoods. The results show that large networks can be effectively controlled at a cost that grows only linearly with network size.
Epstein {\it et al.}\cite{epstein2024strong}
show that symmetric networks of identical Koper oscillators can exhibit strong symmetry-breaking rhythms, where each oscillator displays a distinct oscillation type. Key folded nodes govern the dynamics, and maximal canards in phase space separate the resulting behaviors.
Ocampo-Espindola {\it et al.}\cite{luis2024optimal} study the entrainment of electrochemical oscillators, both numerically and experimentally, using suitably designed external waveforms to induce target phase patterns. Addressing oscillator heterogeneities through phase response curves, they demonstrate that tailored global signals can achieve desired synchronization behaviors---an approach relevant to both engineered and biological systems.
Lee {\it et al.}\cite{lee2024complexified} return to the classic Kuramoto model, generalizing both the parameters and the state variables to the complex domain and identifying a number of intriguing synchronization behaviors in the resulting systems. 
Bunimovich and Skums\cite{bunimovich2024fractal} present a framework for rigorously analyzing self-similarity and fractality in complex networks. They introduce graph-theoretical analogs of Lebesgue and Hausdorff dimensions, linking these to structural features such as community overlaps and graph colorings. 
Finally, 
Brooks and Porter\cite{brooks2025opinion} develop a new model of opinion dynamics on social media that incorporates the concept of bounded confidence, where agents adjust their opinions through interaction. They also explore a notion of opinion reproduction number to quantify the spread of opinions within the network.

A network acting as a reservoir computer is combined with a next-generation dynamical reconstruction in the paper by Chepuri {\it et al.}\cite{chepuri2024hybridizing}. This avoids the necessity of larger, complex reservoir computers for high-dimensional dynamical systems analysis. Each dynamical approach (standard reservoir computers and next generation reservoirs) appears to address the other's shortcomings, thus improving the accuracy and avoiding the need for imposingly large networks. 
The work by Bertalan {\it et al.}\cite{bertalan2024transformations} constructs transformations to establish functional equivalence between neural networks with different architectures. The framework is demonstrated on tasks involving scalar functions, vector fields, and image data, with connections to transfer learning.
Chalkiadakis {\it et al.}\cite{chalkiadakis2023quantum} employ quantum neural networks (QNNs) to design and implement quantum error-correcting codes. Their modified QNNs successfully correct bit-flip and amplitude-damping errors, outperforming standard models. This approach offers a promising pathway for developing adaptive, channel-specific error correction in quantum computing.

A number of other papers in this issue focus on quantum systems. Cuzzuol and Montorsi\cite{cuzzuol2024fundamental} show that hidden properties, captured by nonlocal order parameters, are crucial for identifying different quantum phases in the one-dimensional extended Bose–Hubbard model. Their results suggest accessible experimental methods to detect these phases through local density measurements. In a comparative study of the fundamental features of the transverse field Ising model and the double-well equation, Thudiyangal {\it et al.}\cite{thudiyangal2024nonlinear} take a step towards bridging the gap between information processing in quantum linear systems and classic nonlinear systems. Rangi {\it et al.}\cite{rangi2024out} analyze the out-of-time-order correlator (OTOC) in the disordered Hubbard model using non-equilibrium dynamical mean-field theory and coherent potential approximation. They find regimes for which random local disorder accelerates OTOC decay in the metallic phase, indicating enhanced quantum information scrambling in correlated electron systems. Mazumdar and Clay\cite{mazumdar2024computational} offer a review of Cooper pairs and paired electron ``liquids'' in the quarter-filled bands. This involves superconducting states in cuprates, which have still to be completely explained by quantum theory. Carmelo and Sacramento\cite{carmelo2024ising} review research on the spin-1/2 XXZ chain in the Ising regime, emphasizing that conventional spinon-based representations are insufficient. By employing a physical-spin representation valid in the thermodynamic limit, they describe new quantum transport behaviors and excitation dynamics beyond the traditional spinon paradigm.

Papers on materials-related problems in nonlinear science include the contribution from Hentschel and Procaccia\cite{hentschel2024elastic}, who examine the elastic-to-plastic transition in amorphous solids, revealing that even minimal strain induces localized, quadrupolar plastic events. These events disrupt classical elasticity, necessitating revised models to accurately describe deformation in disordered materials.  Andersen and Kenkre\cite{andersen2024surprising} explore transitions in a two-site quantum system with a cubic self-interaction, identifying and characterizing some novel effects that appear as parameters are varied. Savin and Kivshar\cite{savin2024stabilization} model the formation of helical chains in carbon nanotubes, which support the formation of robust zigzag structures along which proton transport can occur. Mathis {\it et al.}\cite{mathis2024self} revisit the AlChemy model---an artificial chemistry model based on the $\lambda$-calculus---to explore how simple computational rules can lead to complex self-organizing structures. Their analysis reveals that stable organizations emerge more frequently than previously thought, offering insights into both computational applications and the origins of life.

A variety of problems in chaos and fractals are also treated in this Focus Issue. McDonough and Herczyński\cite{mcdonough2024fractal} show that generalizing fractal dimensions to multiple sets and characteristics improves the analysis of complex images, particularly for neural network-based methods. Wei {\it et al.}\cite{wei2024analysis} analyze the complex problem of ship response and potential capsizing under irregular wave forcing. By quantifying each ship’s dynamic behavior under such conditions, their approach offers guidance for operation in random sea states. Danieli {\it et al.}\cite{danieli2024dynamical} use split-step integration to replace integrable dynamics by chaotic ones, showing that the chaotic dynamics in the Toda chain can lead to severe breakdown of the numerical integration. Surprisingly, this helps explain the origin of errors in certain quantum computations. Viana {\it et al.}\cite{viana2024fractal} investigate how passive scalars behave in time-dependent, two-dimensional fluid flows with open boundaries. They describe fractal escape basins and Wada basin boundaries---where a single point borders three or more outcomes---highlighting the complex, chaotic nature of passive scalar transport in open hydrodynamic systems. 
Pikovsky \cite{pikovsky2023deterministic} explains how active particles, having an internal energy source that controls their speed, move at constant speed, and thus external forces affect only their direction. For identical particles, the dynamics are conservative, akin to Hamiltonian optics, and generally exhibit both chaotic and regular regions. 

Localized states---one of the areas in which Campbell’s contributions run the deepest---are treated in a half dozen papers.
Ostrovsky \textit{et al.} \cite{ostrovsky2024localized} demonstrate that waves beyond solitons can arise from a broad class of localized structures---particularly radiating solitons observed in oceanic gravity waves. They extend these findings to systems with non-standard nonlinearities by employing non-analytic functions, offering insights into the origin and formation of rogue waves.
Malomed\cite{malomed2024basic} reviews fractional nonlinear wave propagation models based on Riesz derivatives, originating from quantum mechanics and optical systems. He discusses soliton solutions with cubic or quadratic nonlinearities and highlights experimental realization of fractional group-velocity dispersion in fiber lasers. Cooper {\it et al.}\cite{cooper2024application} construct confining potentials that render $N$-soliton waveforms exact solutions of the Gross-Pitaevskii equation, which models the ground state of bosonic systems. Their approach addresses both attractive and repulsive interactions, providing insights into the controlled formation of soliton structures in Bose–Einstein condensates. Rosenau and Krylov\cite{rosenau2023solitary} analyze microelectromechanical lattices modeled by nonlinear Klein–Gordon equations. They identify stable solitary waveforms, including sharp-fronted ``mesons'' and flat-top ``flatons,'' and introduce a stabilization algorithm enabling persistent soliton propagation in systems where basic solitons are typically unstable. Martin-Vergara {\it et al.}\cite{martin2023discrete} introduce a deflation-based numerical method to uncover discrete breathers---localized, time-periodic excitations---in nonlinear Klein-Gordon lattices. Their approach efficiently identifies diverse multibreather solutions without requiring prior knowledge of their spatial profiles. Ronetti {\it et al.}\cite{ronetti2024levitons} review how single-electron excitations, known as levitons, behave in strongly correlated nanoscale systems. They explore leviton interactions in fractional quantum Hall and superconducting regimes, highlighting phenomena like fractional charge tunneling and energy-entangled electron states. These insights advance quantum transport understanding and have potential applications in quantum information.

A final cluster of papers addresses problems in living systems. Speakman and Gunaratne\cite{speakman2024kneading} propose a novel mathematical framework for RNA splicing by mapping nucleotide sites into a plane defined by their surrounding sequences. This approach aims to uncover underlying rules of splicing and identify potential sources of errors, offering insights into gene expression and associated diseases. 
Kuzmin \textit{et al.} \cite{kuzmin2024dynamics} examine the evolution of genes affected by copying errors and how cellular repair mechanisms can introduce genetic changes that occasionally remain uncorrected. This process accelerates karyotype changes (large chromosomal alterations) that would otherwise occur over tens of thousands of years, and in some cases, contributes to cancer development.
Nelias {\it et al.}\cite{nelias2024tapping} investigate how professional drummers regulate tapping strength during rhythmic tasks, expanding on previous studies focused on the control of timing in rhythmic movement. Their analysis of time-series data shows that tapping intensity exhibits $1/f^\beta$ noise and long-range correlations, indicating scale-free fluctuations. 
Peyrard \cite{peyrard2023can} analyzes the temporal evolution of COVID-19, revealing quasi-periodic, structure-like recurrences. Understanding the origin of these recurrences is a central part of the study, which considers several potential mechanisms---one of which accurately reproduces the observed dynamics.

\section{Outlook}

The papers in this Focus Issue reflect both the sustained strength of foundational topics in nonlinear dynamics and the accelerating momentum of emerging directions. Traditional areas such as chaos, fractals, and localized states continue to yield deep insights, while rapid advances in network science are opening new conceptual and methodological frontiers. As nonlinear science increasingly intersects with fields like machine learning, quantum dynamics, materials science, and living systems, we anticipate that these vibrant and evolving areas will drive new theoretical developments and innovative applications in the years ahead. Together, the contributions in this Focus Issue illustrate how nonlinear science is expanding its reach---offering powerful tools to understand complexity across physical, biological, and technological domains.

This rich, far-reaching collection of work is a fitting tribute to David K. Campbell, who is ``... surely one of a small number of investigators worldwide [who] have brought the field of nonlinear science to its present status'' \cite{andersen2024surprising}. ``Through many decades of work and inspiration,'' as noted in Ref.\ [\citenum{lee2024complexified}], ``... he has created an enduring legacy that overarches the full range from observed nonlinear phenomena in nature to advancing mathematical tools for their analysis.''
In his own work, and in his shepherding of the work of others and the growth of the field of  nonlinear science, ``... David once again did things in the way that they are supposed to be done.'' \cite{brooks2025opinion}. 

The editors of this Focus Issue want to express their personal thanks and appreciation to Campbell for all of this work and their best wishes for many productive and happy years to come.

\begin{acknowledgments}
We extend our sincere gratitude to all the authors of this landmark Focus Issue for their invaluable and inspiring contributions. We also thank the staff and editor-in-chief of {\it Chaos}, Prof.\ J\"urgen Kurths, for their indispensable support.
\end{acknowledgments}

\bigskip
\noindent
{\bf Credits:}
This article may be downloaded for personal use only. Any other use requires prior permission of the author(s) and AIP Publishing. This article appeared in Chaos 35, 070402 (2025) and may be found at  \url{https://doi.org/10.1063/5.0281491}.

\appendix

\section*{Appendix: Kenneth Ford's Historical Account}

The story of the journal {\it Chaos} begins, as so many stories do, with a consideration of money.  In the late 1980s, my organization, AIP, was translating into English and selling twenty Soviet physics journals under a single umbrella contract with the Soviet Copyright Agency.   But this was when the Soviet Union was coming unglued. 

To help us negotiate this new, rocky landscape, we retained the services of Martin Levin, a New York attorney with extensive experience in the USSR.  One of his early pieces of advice was, “You should offer the editors what competing commercial publishers cannot offer: science.” This led, quickly, after discussions on both sides of the Iron Curtain, to the idea of joint Soviet-American conferences held in the two countries in alternate years, jointly sponsored by AIP and the Soviet Academy of Sciences. But on what subject?
 
I think it was we at AIP who suggested, as criteria, that the subject be primarily theoretical, not have too many practitioners, be currently very active, and be of comparable strength in the two countries. Somehow, nonlinear science, and chaos in particular, emerged as a suitable topic. In the USSR, Roald Sagdeev and George Zaslavsky agreed to anchor their end of a planning group. In this country, the name David Campbell kept coming up as the logical candidate for the western anchor.
 
Fortunately, David said yes to leading the US end of the Soviet-American Chaos/XAOC Conferences, and I have to describe them as a smashing success. The first, third, and fifth conferences---in 1989, 1991, and 1993---were held at the National Academy of Sciences Conference Center in Woods Hole, Massachusetts; the second---in 1990---in Tarusa, Russia; and the fourth---in 1992---in Kyiv, Ukraine. 
 
Serious discussion of a possible new journal on nonlinear science, possibly with an emphasis on chaos, occurred as early as the summer of 1989 at the first Chaos/XAOC conference in Woods Hole.  It is something of a miracle that less than a full year passed before I was able to write to David, inviting him to be the Editor of the newly approved {\it Chaos} journal. In that letter, I called him the people’s choice, which indeed he was. How fortunate we all are that David accepted and guided {\it Chaos} to be the widely respected international journal that it is today.

What a pleasure it was, David, to work with you as {\it Chaos} was taking shape. What a pleasure it has been since then to watch this journal, under your leadership, grow and prosper.

%

\providecommand{\noopsort}[1]{}\providecommand{\singleletter}[1]{#1}%

\end{document}